\documentclass[aps,preprint]{revtex4}%
\usepackage{amsfonts}
\usepackage{amsmath}
\usepackage{amssymb}
\usepackage{graphicx}%
\begin{document}
\preprint{ }
\title{The ground state energy at unitarity}
\author{Dean Lee}
\affiliation{Department of Physics, North Carolina State University, Raleigh, NC 27695}

\begin{abstract}
We consider two-component fermions on the lattice in the unitarity limit.
\ This is an idealized limit of attractive fermions where the range of the
interaction is zero and the scattering length is infinite. \ Using Euclidean
time projection, we compute the ground state energy using four computationally
different but physically identical auxiliary-field methods. \ The best
performance is obtained using a bounded continuous auxiliary field and a
non-local updating algorithm called hybrid Monte Carlo. \ With this method we
calculate results for $10$ and $14$ fermions at lattice volumes $4^{3}%
,5^{3},6^{3},7^{3},8^{3}$ and extrapolate to the continuum limit. \ For $10$
fermions in a periodic cube, the ground state energy is $0.292(12)$ times the
ground state energy for non-interacting fermions. \ For $14$ fermions the
ratio is $0.329(5)$.

\end{abstract}
\maketitle

\section{Introduction}

The unitarity limit describes attractive two-component fermions in an
idealized limit where the range of the interaction is zero and the scattering
length is infinite. \ The name refers to the fact that the S-wave cross
section saturates the limit imposed by unitarity, $\sigma_{0}(k)\leq4\pi
/k^{2}$, for low momenta $k$. \ In the unitarity limit details about the
microscopic interaction are lost, and the system displays universal
properties. \ Throughout our discussion we refer to the two degenerate
components as up and down spins, though the correspondence with actual spin is
not necessary. \ At sufficiently low temperatures the spin-unpolarized system
is believed to be superfluid with properties in between a
Bardeen-Cooper-Schrieffer (BCS) fermionic superfluid at weak coupling and a
Bose-Einstein condensate of dimers at strong coupling
\cite{Eagles:1969PR,Leggett:1980pro,Nozieres:1985JLTP}.

In nuclear physics the phenomenology of the unitarity limit is relevant to
cold dilute neutron matter. \ The scattering length for elastic
neutron-neutron collisions is about $-18.5$ fm while the range of the
interaction is comparable to the Compton wavelength of the pion, $m_{\pi}%
^{-1}\approx1.4$ fm. \ Therefore the unitarity limit is approximately realized
when the interparticle spacing is about $5$ fm. \ While these conditions
cannot be produced experimentally, neutrons at around this density may be
present in the inner crust of neutron stars
\cite{Pethick:1995di,Lattimer:2004pg}.

The unitarity limit has been experimentally realized with cold atomic Fermi
gases. For alkali atoms the relevant length scale for the interatomic
potential at long distances is the van der Waals length $\ell_{\text{vdW}}$.
\ If the spacing between atoms is much larger than $\ell_{\text{vdW}}$, then
to a good approximation the interatomic potential is equivalent to a zero
range interaction. \ The scattering length can be manipulated using a magnetic
Feshbach resonance
\cite{O'Hara:2002,Gupta:2002,Regal:2003,Bourdel:2003,Gehm:2003,Bartenstein:2004,Kinast:2005}%
. \ This technique involves a molecular bound state in a \textquotedblleft
closed\textquotedblright\ hyperfine channel crossing the scattering threshold
of an \textquotedblleft open\textquotedblright\ channel. \ The magnetic
moments for the two channels are typically different, and so the crossing can
be produced using an applied magnetic field.

At zero temperature there are no dimensionful parameters other than particle
density. \ For $N_{\uparrow}$ up spins and $N_{\downarrow}$ down spins in a
given volume we write the energy of the unitarity-limit ground state as
$E_{N_{\uparrow},N_{\downarrow}}^{0}$. \ For the same volume we call the
energy of the free non-interacting ground state $E_{N_{\uparrow}%
,N_{\downarrow}}^{0\text{,free}}$, and the dimensionless ratio of the two
$\xi_{N_{\uparrow},N_{\downarrow}}$,%
\begin{equation}
\xi_{N_{\uparrow},N_{\downarrow}}=E_{N_{\uparrow},N_{\downarrow}}%
^{0}/E_{N_{\uparrow},N_{\downarrow}}^{0\text{,free}}.
\end{equation}
The parameter $\xi$ is defined as the thermodynamic limit for the
spin-unpolarized system,%
\begin{equation}
\xi=\lim_{N\rightarrow\infty}\xi_{N,N}.
\end{equation}
Several experiments have measured $\xi$ from the expansion of $^{6}$Li and
$^{40}$K released from a harmonic trap. \ Some recent measured values for
$\xi$ are $0.51(4)$ \cite{Kinast:2005}, $0.46_{-05}^{+12}$ \cite{Stewart:2006}%
, and $0.32_{-13}^{+10}$ \cite{Bartenstein:2004}. \ The disagreement between
these measurements and larger values for $\xi$ reported in earlier experiments
\cite{O'Hara:2002,Bourdel:2003,Gehm:2003} suggest further work may be needed.

There are numerous analytic calculations of $\xi$ using a variety of
techniques such as BCS saddle point and variational approximations, Pad\'{e}
approximations, mean field theory, density functional theory, exact
renormalization group, dimensional $\epsilon$-expansions, and large-$N$
expansions
\cite{Engelbrecht:1997,Baker:1999dg,Heiselberg:1999,Perali:2004,Schafer:2005kg,Papenbrock:2005,Nishida:2006a,Nishida:2006b,JChen:2006,Krippa:2007A,Arnold:2007,Nikolic:2007,Veillette:2006}%
. \ The values for $\xi$ range from $0.2$ to $0.6$. \ Fixed-node Green's
function Monte Carlo simulations for a periodic cube have found $\xi_{N,N}$ to
be $0.44(1)$ for $5\leq N\leq21$ \cite{Carlson:2003z} and $0.42(1)$ for larger
$N$ \cite{Astrakharchik:2004,Carlson:2005xy}. \ A restricted path integral
Monte Carlo calculation finds similar results \cite{Akkineni:2006A}, and a
sign-restricted mean field lattice calculation yields $0.449(9)$
\cite{Juillet:2007a}. \ 

There have also been simulations of two-component fermions on the lattice in
the unitarity limit at nonzero temperature. \ When data are extrapolated to
zero temperature the results of \cite{Bulgac:2005a,Bulgac:2008b} produce a
value for $\xi$ similar to the fixed-node results. \ The same is true for
\cite{Burovski:2006a,Burovski:2006b}, though with significant error bars.
\ The extrapolated zero temperature lattice results from
\cite{Lee:2005is,Lee:2005it} established a bound, $0.07\leq\xi\leq0.42$, while
more recent lattice calculations yield $0.261(12)$
\cite{Abe:2007fe,Abe:2007ff}. \ The work of \cite{Abe:2007fe,Abe:2007ff}
includes a comparison of two lattice algorithms similar to the ones discussed
in this paper.

In \cite{Lee:2005fk}\ $\xi_{N,N}$ was calculated on the lattice using
Euclidean time projection for $N=3,5,7,9,11$ and lattice volumes $4^{3}%
,5^{3},6^{3}$. \ From these small volumes it was estimated that $\xi=0.25(3)$.
\ More recent results using a technique called the symmetric heavy-light
ansatz found similar values for $\xi_{N,N}$ at the same lattice volumes and
estimated $\xi=0.31(1)$ in the continuum and thermodynamic limits
\cite{Lee:2007A}.

For lattice calculations of $\xi_{N,N}$ at fixed $N$, systematic errors due to
nonzero lattice spacing can be removed by extrapolating to the continuum
limit. \ At unitarity where the scattering length is set to infinity, this
continuum extrapolation corresponds with increasing the lattice volume as
measured in lattice units. \ In this paper we present new lattice results
using the Euclidean time projection method introduced in \cite{Lee:2005fk} and
extrapolate to the continuum limit.

In earlier work \cite{Lee:2005fk,Lee:2006hr} singular matrices were
encountered that prevented calculations at larger volumes. \ We discuss this
and related issues by considering four different auxiliary-field methods which
reproduce exactly the same Euclidean lattice model. \ These four methods
include a Gaussian-integral auxiliary field, an exponentially-coupled
auxiliary field, a discrete auxiliary field, and a bounded continuous
auxiliary field. \ By far the best performance is obtained using the bounded
continuous auxiliary field and hybrid Monte Carlo \cite{Duane:1987de}. \ With
this method we calculate the ground state energy for $10$ and $14$ fermions at
lattice volumes $4^{3},5^{3},6^{3},7^{3},8^{3}$ and extrapolate to the
continuum limit.

\section{Euclidean time lattice formalism}

Throughout our discussion of the lattice formalism we use dimensionless
parameters and operators corresponding with physical values multiplied by the
appropriate power of the spatial lattice spacing $a$. \ In our notation the
three-component integer vector $\vec{n}$ labels the lattice sites of a
three-dimensional periodic lattice with dimensions $L^{3}$. \ The spatial
lattice unit vectors are denoted $\hat{l}=\hat{1}$, $\hat{2}$, $\hat{3}$. \ We
use $n_{t}$ to label lattice steps in the temporal direction. $\ L_{t}$
denotes the total number of lattice time steps. \ The temporal lattice spacing
is given by $a_{t}$, and $\alpha_{t}=a_{t}/a$ is the ratio of the temporal to
spatial lattice spacing. \ We also define $h=\alpha_{t}/(2m)$, where $m$ is
the fermion mass in lattice units.

\begin{figure}
[ptb]
\begin{center}
\includegraphics[
height=3.2915in,
width=5.1949in
]%
{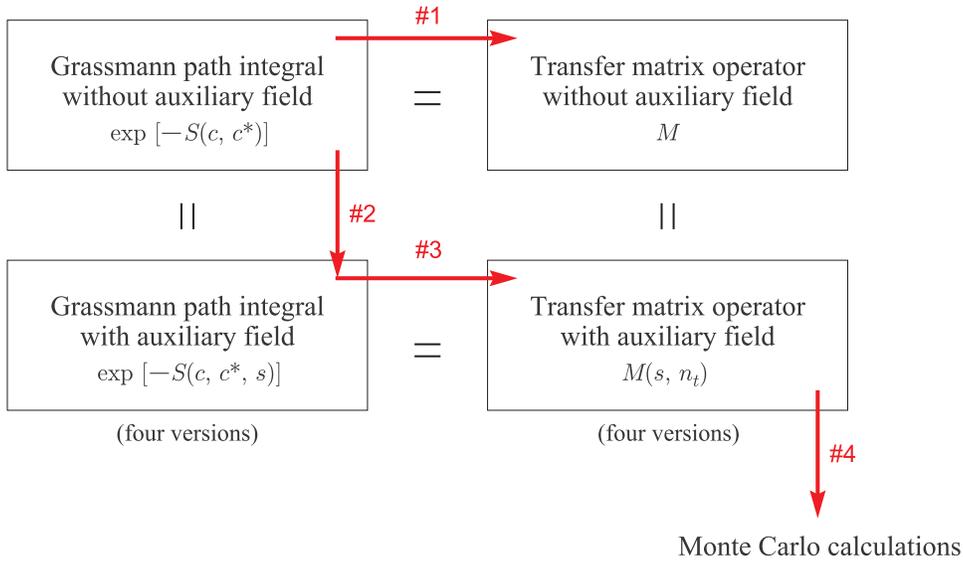}%
\caption{(Color online) Schematic diagram showing four equivalent Euclidean
time lattice formulations. \ The numbered arrows indicate the order of
discussion in the text.}%
\label{formalisms}%
\end{center}
\end{figure}

We discuss four different formulations of the Euclidean time lattice for
interacting fermions: \ Grassmann path integrals with and without auxiliary
field and transfer matrix operators with and without auxiliary field. \ For
any spatial and temporal lattice spacings these four formulations produce
exactly the same physics, as indicated in Fig. \ref{formalisms}. \ We follow
the order of discussion indicated by the numbered arrows in Fig.
\ref{formalisms}.%

\subsection{Grassmann path integral without auxiliary field}

Let $c_{i}$ and $c_{i}^{\ast}$ be anticommuting Grassmann fields for spin $i$.
\ The Grassmann fields are periodic with respect to the spatial lengths of the
$L^{3}$ lattice,%
\begin{equation}
c_{i}(\vec{n}+L\hat{1},n_{t})=c_{i}(\vec{n}+L\hat{2},n_{t})=c_{i}(\vec
{n}+L\hat{3},n_{t})=c_{i}(\vec{n},n_{t}),
\end{equation}
and antiperiodic along the temporal direction,%
\begin{equation}
c_{i}(\vec{n},n_{t}+L_{t})=-c_{i}(\vec{n},n_{t}).
\end{equation}
We write $DcDc^{\ast}$ as shorthand for the integral measure,%
\begin{equation}
DcDc^{\ast}=\prod_{\vec{n},n_{t},i=\uparrow,\downarrow}dc_{i}(\vec{n}%
,n_{t})dc_{i}^{\ast}(\vec{n},n_{t}).
\end{equation}
The Grassmann spin densities, $\rho_{\uparrow}$ and $\rho_{\downarrow},$ are
defined as%
\begin{equation}
\rho_{\uparrow}(\vec{n},n_{t})=c_{\uparrow}^{\ast}(\vec{n},n_{t})c_{\uparrow
}(\vec{n},n_{t}),
\end{equation}%
\begin{equation}
\rho_{\downarrow}(\vec{n},n_{t})=c_{\downarrow}^{\ast}(\vec{n},n_{t}%
)c_{\downarrow}(\vec{n},n_{t}).
\end{equation}
We consider the Grassmann path integral
\begin{equation}
\mathcal{Z}=\int DcDc^{\ast}\exp\left[  -S\left(  c,c^{\ast}\right)  \right]
,
\end{equation}%
\begin{equation}
S(c,c^{\ast})=S_{\text{free}}(c,c^{\ast})+C\alpha_{t}\sum_{\vec{n},n_{t}}%
\rho_{\uparrow}(\vec{n},n_{t})\rho_{\downarrow}(\vec{n},n_{t}).
\label{path_nonaux}%
\end{equation}
The action $S(c,c^{\ast})$ consists of the free fermion action%
\begin{align}
S_{\text{free}}(c,c^{\ast})  &  =\sum_{\vec{n},n_{t},i=\uparrow,\downarrow
}\left[  c_{i}^{\ast}(\vec{n},n_{t})c_{i}(\vec{n},n_{t}+1)-(1-6h)c_{i}^{\ast
}(\vec{n},n_{t})c_{i}(\vec{n},n_{t})\right] \nonumber\\
&  -h\sum_{\vec{n},n_{t},i=\uparrow,\downarrow}\sum_{l=1,2,3}\left[
c_{i}^{\ast}(\vec{n},n_{t})c_{i}(\vec{n}+\hat{l},n_{t})+c_{i}^{\ast}(\vec
{n},n_{t})c_{i}(\vec{n}-\hat{l},n_{t})\right]  ,
\end{align}
and an attractive contact interaction between up and down spins. \ We take
this action as the definition of the lattice model. \ All other formulations
introduced later will be shown to be exactly equivalent.

We use a fermion mass of $939$ MeV and lattice spacings $a=(50$ MeV$)^{-1}$,
$a_{t}=(24$ MeV$)^{-1}$.\ \ In dimensionless lattice units the corresponding
parameters are $m=18.78$ and $\alpha_{t}=2.0833$. \ These are the same values
used in \cite{Lee:2005fk} and is motived by the relevant physical scales for
dilute neutron matter.

\subsection{Interaction coefficient $C$ for general spatial and temporal
lattice spacings}

We use L\"{u}scher's formula
\cite{Luscher:1986pf,Beane:2003da,Seki:2005ns,Borasoy:2006qn} to relate the
coefficient $C$ to the S-wave scattering length. \ We consider one up-spin
particle and one down-spin particle in a periodic cube of length $L$.
\ L\"{u}scher's formula relates the two-particle energy levels in the
center-of-mass frame to the S-wave phase shift,%
\begin{equation}
p\cot\delta_{0}(p)=\frac{1}{\pi L}S\left(  \eta\right)  ,\qquad\eta=\left(
\frac{Lp}{2\pi}\right)  ^{2}, \label{lusch}%
\end{equation}
where $S(\eta)$ is the three-dimensional zeta function,%
\begin{equation}
S(\eta)=\lim_{\Lambda\rightarrow\infty}\left[  \sum_{\vec{n}}\frac
{\theta(\Lambda^{2}-\vec{n}^{2})}{\vec{n}^{2}-\eta}-4\pi\Lambda\right]  .
\label{S}%
\end{equation}
For small momenta the effective range expansion gives%
\begin{equation}
p\cot\delta_{0}(p)\approx-\frac{1}{a_{\text{scatt}}}+\frac{1}{2}r_{0}%
p^{2}+\cdots\text{,} \label{effrange}%
\end{equation}
where $a_{\text{scatt}}$ is the scattering length and $r_{0}$ is the effective range.

\begin{figure}
[ptb]
\begin{center}
\includegraphics[
height=0.9617in,
width=4.0041in
]%
{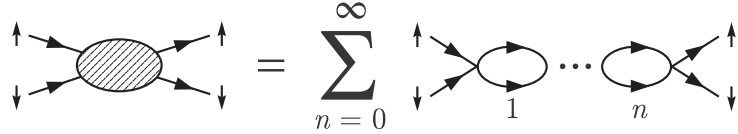}%
\caption{Sum of bubble diagrams contributing to two-particle scattering.}%
\label{twotwo}%
\end{center}
\end{figure}

In terms of $\eta$, the energy of the two-particle scattering state is%
\begin{equation}
E_{\text{pole}}=\frac{p^{2}}{m}=\frac{\eta}{m}\left(  \frac{2\pi}{L}\right)
^{2}. \label{Epole}%
\end{equation}
We compute the location of the two-particle scattering pole by summing the
bubble diagrams shown in Fig.~\ref{twotwo}.%
After writing the sum of bubble diagrams as a geometric series, the relation
between $C$ and $E_{\text{pole}}$ becomes
\begin{equation}
-\frac{1}{C\alpha_{t}}=\lim_{L\rightarrow\infty}\frac{1}{L^{3}}\sum_{\vec
{k}\text{ }\operatorname{integer}}\frac{1}{e^{-E_{\text{pole}}\alpha_{t}%
}-1+2\alpha_{t}\omega(2\pi\vec{k}/L)-\alpha_{t}^{2}\omega^{2}(2\pi\vec{k}/L)},
\label{bubble}%
\end{equation}
where%
\begin{equation}
\omega(\vec{p})=\frac{1}{m}\sum_{l=1,2,3}\left(  1-\cos p_{l}\right)  .
\end{equation}
These expressions were first derived in \cite{Lee:2004qd}. \ In the unitarity
limit the scattering length $a_{\text{scatt}}$ is set to infinity and all
other effective range parameters are set to zero. \ From Eq.~(\ref{effrange})
and (\ref{lusch}) this is equivalent to setting $S\left(  \eta\right)  =0$.
\ To minimize lattice discretization errors we pick the root of $S\left(
\eta\right)  $ closest to zero, $\eta=-0.095901$. \ After selecting some large
value for $L$, we use Eq.~(\ref{Epole}) and (\ref{bubble}) to determine the
unitarity limit value for the coefficient $C$. \ For our chosen lattice
parameters $C$ is $-7.4415\times10^{-5}$ MeV$^{-2}$ in physical units or
$-0.18604$ in dimensionless lattice units.

\subsection{Transfer matrix operator without auxiliary field}

We convert the Grassmann path integral into the trace of a product of transfer
matrix operators using the exact correspondence
\cite{Creutz:1988wv,Creutz:1999zy}%
\begin{align}
&  Tr\left\{  \colon F_{L_{t}-1}\left[  a_{i^{\prime}}^{\dagger}(\vec
{n}^{\prime}),a_{i}(\vec{n})\right]  \colon\cdot\cdots\cdot\colon F_{0}\left[
a_{i^{\prime}}^{\dagger}(\vec{n}^{\prime}),a_{i}(\vec{n})\right]
\colon\right\} \nonumber\\
&  =\int DcDc^{\ast}\exp\left\{  \sum_{n_{t}=0}^{L_{t}-1}\sum_{\vec{n},i}%
c_{i}^{\ast}(\vec{n},n_{t})\left[  c_{i}(\vec{n},n_{t})-c_{i}(\vec{n}%
,n_{t}+1)\right]  \right\} \nonumber\\
&  \qquad\qquad\qquad\times\prod_{n_{t}=0}^{L_{t}-1}F_{n_{t}}\left[
c_{i^{\prime}}^{\ast}(\vec{n}^{\prime},n_{t}),c_{i}(\vec{n},n_{t})\right]  ,
\label{correspondence}%
\end{align}
where $c_{i}(\vec{n},L_{t})=-c_{i}(\vec{n},0).$ \ We use $a_{i}(\vec{n})$ and
$a_{i}^{\dagger}(\vec{n})$ to denote fermion annihilation and creation
operators respectively for spin $i$ at lattice site $\vec{n}$. \ The $\colon$
symbols in (\ref{correspondence}) indicate normal ordering. \ Let us define%
\begin{equation}
H_{\text{free}}=\frac{3}{m}\sum_{\vec{n},i=\uparrow,\downarrow}a_{i}^{\dagger
}(\vec{n})a_{i}(\vec{n})-\frac{1}{2m}\sum_{\vec{n},i=\uparrow,\downarrow}%
\sum_{l=1,2,3}\left[  a_{i}^{\dagger}(\vec{n})a_{i}(\vec{n}+\hat{l}%
)+a_{i}^{\dagger}(\vec{n})a_{i}(\vec{n}-\hat{l})\right]
\end{equation}
and%
\begin{equation}
\rho_{\uparrow}^{a^{\dagger}a}(\vec{n})=a_{\uparrow}^{\dagger}(\vec
{n})a_{\uparrow}(\vec{n}),
\end{equation}%
\begin{equation}
\rho_{\downarrow}^{a^{\dagger}a}(\vec{n})=a_{\downarrow}^{\dagger}(\vec
{n})a_{\downarrow}(\vec{n}).
\end{equation}
Then%
\begin{equation}
\mathcal{Z}=Tr\left(  M^{L_{t}}\right)  ,
\end{equation}
where $M$ is the normal-ordered transfer matrix operator%
\begin{equation}
M=:\exp\left[  -H_{\text{free}}\alpha_{t}-C\alpha_{t}\sum_{\vec{n}}%
\rho_{\uparrow}^{a^{\dagger}a}(\vec{n})\rho_{\downarrow}^{a^{\dagger}a}%
(\vec{n})\right]  :. \label{transfer_noaux}%
\end{equation}

\subsection{Grassmann path integral with auxiliary field}

We consider four different auxiliary-field transformations labelled with the
subscript $j=1,2,3,4$. \ Let us define the Grassmann actions%
\begin{equation}
S_{j}\left(  c,c^{\ast},s\right)  =S_{\text{free}}(c,c^{\ast})-\sum_{\vec
{n},n_{t}}A_{j}\left[  s(\vec{n},n_{t})\right]  \cdot\left[  \rho_{\uparrow
}(\vec{n},n_{t})+\rho_{\downarrow}(\vec{n},n_{t})\right]  , \label{Aj}%
\end{equation}
and Grassmann path integrals%
\begin{equation}
\mathcal{Z}_{j}=\prod\limits_{\vec{n},n_{t}}\left[  \int d_{j}s(\vec{n}%
,n_{t})\right]  \int DcDc^{\ast}\exp\left[  -S_{j}\left(  c,c^{\ast},s\right)
\right]  ,
\end{equation}
where $s$ is a real-valued function over all lattice sites. \ For $j=1$, we
use a Gaussian-integral transformation similar to the original
Hubbard-Stratonovich transformation \cite{Stratonovich:1958,Hubbard:1959ub},
\begin{equation}
\int d_{1}s(\vec{n},n_{t})=\frac{1}{\sqrt{2\pi}}\int_{-\infty}^{+\infty
}ds(\vec{n},n_{t})e^{-\frac{1}{2}s^{2}(\vec{n},n_{t})},
\end{equation}%
\begin{equation}
A_{1}\left[  s(\vec{n},n_{t})\right]  =\sqrt{-C_{1}\alpha_{t}}\,s(\vec
{n},n_{t}).
\end{equation}
For $j=2$ we use an auxiliary field with exponential coupling to the fermion
densities,%
\begin{equation}
\int d_{2}s(\vec{n},n_{t})=\frac{1}{\sqrt{2\pi}}\int_{-\infty}^{+\infty
}ds(\vec{n},n_{t})e^{-\frac{1}{2}s^{2}(\vec{n},n_{t})},
\end{equation}%
\begin{equation}
A_{2}\left[  s(\vec{n},n_{t})\right]  =(1-6h)\left[  e^{\sqrt{-C_{2}\alpha
_{t}}s(\vec{n},n_{t})+\frac{C_{2}\alpha_{t}}{2}}-1\right]  .
\end{equation}
This is the auxiliary-field transformation used in
\cite{Lee:2004qd,Lee:2005is,Lee:2005it,Lee:2005fk,Lee:2006hr}. \ For $j=3$ we
use a discrete auxiliary field taking values $-1$ or $+1$ at each lattice
site,%
\begin{equation}
\int d_{3}s(\vec{n},n_{t})=\frac{1}{2}\sum_{s(\vec{n},n_{t})=\pm1},
\end{equation}%
\begin{equation}
A_{3}\left[  s(\vec{n},n_{t})\right]  =\sqrt{-C_{3}\alpha_{t}}\,s(\vec
{n},n_{t}).
\end{equation}
The discrete Hubbard-Stratonovich transformation was first introduced by
Hirsch in \cite{Hirsch:1983} and has been used in a number of recent lattice
simulations \cite{Bulgac:2005a,Lee:2005xy,Abe:2007fe,Abe:2007ff,Bulgac:2008b}.
\ For $j=4$ we introduce a new transformation which uses a bounded but
continuous auxiliary field,%
\begin{equation}
\int d_{4}s(\vec{n},n_{t})=\frac{1}{2\pi}\int_{-\pi}^{+\pi}ds(\vec{n},n_{t}),
\end{equation}%
\begin{equation}
A_{4}\left[  s(\vec{n},n_{t})\right]  =\sqrt{-C_{4}\alpha_{t}}\sin\left[
s(\vec{n},n_{t})\right]  .
\end{equation}
The motivation for the new transformation is discussed later when comparing
and analyzing computational performance.

For each $j=1,2,3,4$ the integral transformations are defined so that
\begin{equation}
\int d_{j}s(\vec{n},n_{t})1=1,
\end{equation}%
\begin{equation}
\int d_{j}s(\vec{n},n_{t})\,A_{j}\left[  s(\vec{n},n_{t})\right]  =0.
\end{equation}
Since all even products of Grassmann variables commute, we can factor out the
term in $\mathcal{Z}_{j}$ involving the auxiliary field $s$ at $\vec{n},n_{t}%
$. \ To shorten the notation we temporarily omit writing $\vec{n},n_{t}$
explicitly. \ For each $j=1,2,3,4$ we find%
\begin{align}
\int d_{j}s\exp\left[  \,A_{j}\left(  s\right)  \left(  \rho_{\uparrow}%
+\rho_{\downarrow}\right)  \right]   &  =\int d_{j}s\left[  1+\,A_{j}\left(
s\right)  \left(  \rho_{\uparrow}+\rho_{\downarrow}\right)  +A_{j}^{2}\left(
s\right)  \rho_{\uparrow}\rho_{\downarrow}\right] \nonumber\\
&  =1+\int d_{j}s\,A_{j}^{2}\left(  s\right)  \rho_{\uparrow}\rho_{\downarrow
}=\exp\left[  \int d_{j}s\,A_{j}^{2}\left(  s\right)  \rho_{\uparrow}%
\rho_{\downarrow}\right]  .
\end{align}
For the four cases the integral of $A_{j}^{2}$ is%
\begin{equation}
\int d_{1}s\,A_{1}^{2}\left(  s\right)  =-C_{1}\alpha_{t},
\end{equation}%
\begin{equation}
\int d_{2}s\,A_{2}^{2}\left(  s\right)  =(1-6h)^{2}\left(  e^{-C_{2}\alpha
_{t}}-1\right)  ,
\end{equation}%
\begin{equation}
\int d_{3}s\,A_{3}^{2}\left(  s\right)  =-C_{3}\alpha_{t},
\end{equation}%
\begin{equation}
\int d_{4}s\,A_{4}^{2}\left(  s\right)  =-\frac{1}{2}C_{4}\alpha_{t}.
\end{equation}
Therefore%
\begin{equation}
\mathcal{Z}=\mathcal{Z}_{1}=\mathcal{Z}_{2}=\mathcal{Z}_{3}=\mathcal{Z}_{4}
\label{z_equality}%
\end{equation}
when the interaction coefficients satisfy%
\begin{equation}
C=C_{1}=-\frac{(1-6h)^{2}\left(  e^{-C_{2}\alpha_{t}}-1\right)  }{\alpha_{t}%
}=C_{3}=\frac{1}{2}C_{4}.
\end{equation}

\section{Transfer matrix operator with auxiliary field}

Using Eq.~(\ref{correspondence}) and (\ref{z_equality}) we can write
$\mathcal{Z}$ as a product of transfer matrix operators which depend on the
auxiliary fields,%
\begin{equation}
\mathcal{Z}=\prod\limits_{\vec{n},n_{t}}\left[  \int d_{j}s(\vec{n}%
,n_{t})\right]  Tr\left\{  M_{j}(s,L_{t}-1)\cdot\cdots\cdot M_{j}%
(s,0)\right\}  ,
\end{equation}
where%
\begin{equation}
M_{j}(s,n_{t})=\colon\exp\left\{  -H_{\text{free}}\alpha_{t}+\sum_{\vec{n}%
}A_{j}\left[  s(\vec{n},n_{t})\right]  \cdot\left[  \rho_{\uparrow
}^{a^{\dagger}a}(\vec{n})+\rho_{\downarrow}^{a^{\dagger}a}(\vec{n})\right]
\right\}  \colon. \label{transfer_aux}%
\end{equation}
For the numerical lattice calculations presented in this paper we use the
auxiliary-field transfer matrix operators in Eq.~(\ref{transfer_aux}).

Let $\left\vert \Psi_{N,N}^{0,\text{free}}\right\rangle $ be the normalized
Slater-determinant ground state on the lattice for a non-interacting system of
$N$ up spins and $N$ down spins. \ For each spin we fill momentum states in
the order shown in Table \ref{filling}. \ \begin{table}[tb]
\caption{Filling sequence of momentum states for each spin.}%
\label{filling}
\begin{tabular}
[c]{|c|c|}\hline
$N$ & additional momenta filled\\\hline
$1$ & $\left\langle 0,0,0\right\rangle $\\\hline
$3$ & $\left\langle \frac{2\pi}{L},0,0\right\rangle ,\left\langle -\frac{2\pi
}{L},0,0\right\rangle $\\\hline
$5$ & $\left\langle 0,\frac{2\pi}{L},0\right\rangle ,\left\langle
0,-\frac{2\pi}{L},0\right\rangle $\\\hline
$7$ & $\left\langle 0,0,\frac{2\pi}{L}\right\rangle ,\left\langle
0,0,-\frac{2\pi}{L}\right\rangle $\\\hline
\end{tabular}
\end{table}We construct the Euclidean time projection amplitude%
\begin{equation}
Z_{N,N}(t)\equiv\prod\limits_{\vec{n},n_{t}}\left[  \int d_{j}s(\vec{n}%
,n_{t})\right]  \left\langle \Psi_{N,N}^{0,\text{free}}\right\vert
M_{j}(s,L_{t}-1)\cdot\cdots\cdot M_{j}(s,0)\left\vert \Psi_{N,N}%
^{0,\text{free}}\right\rangle ,
\end{equation}
where $t=L_{t}\alpha_{t}$. \ The result upon integration gives the same
$Z_{N,N}(t)$ for each $j=1,2,3,4.$

As a result of normal ordering, $M_{j}(s,n_{t})$ consists of only
single-particle operators interacting with the background auxiliary field and
no direct interactions between particles. \ We find%
\begin{equation}
\left\langle \Psi_{N,N}^{0,\text{free}}\right\vert M_{j}(s,L_{t}-1)\cdot
\cdots\cdot M_{j}(s,0)\left\vert \Psi_{N,N}^{0,\text{free}}\right\rangle
=\left[  \det\mathbf{M}_{j}(s,t)\right]  ^{2}, \label{detsquare}%
\end{equation}
where%
\begin{equation}
\left[  \mathbf{M}_{j}(s,t)\right]  _{k^{\prime}k}=\left\langle \vec
{p}_{k^{\prime}}\right\vert M_{j}(s,L_{t}-1)\cdot\cdots\cdot M_{j}%
(s,0)\left\vert \vec{p}_{k}\right\rangle ,
\end{equation}
for matrix indices $k,k^{\prime}=1,\cdots,N$. $\ \left\vert \vec{p}%
_{k}\right\rangle ,\left\vert \vec{p}_{k^{\prime}}\right\rangle $ are
single-particle momentum states comprising the Slater-determinant
initial/final state. \ The single-particle interactions in $M_{j}(s,n_{t})$
are the same for both up and down spins, and this is why the determinant in
Eq.~(\ref{detsquare}) is squared. \ Since the matrix is real-valued, the
square of the determinant is nonnegative and there is no problem with
oscillating signs.

\bigskip For the interacting lattice system at unitarity, we label the energy
eigenstates $\left\vert \Psi_{N,N}^{k}\right\rangle $ with energies
$E_{N,N}^{k}$ in order of increasing energy,%
\begin{equation}
E_{N,N}^{0}\leq E_{N,N}^{1}\leq\cdots\leq E_{N,N}^{k}\leq\cdots.
\end{equation}
In the transfer matrix formalism these energies are defined in terms of the
logarithm of the transfer matrix eigenvalue,
\begin{equation}
M\left\vert \Psi_{N,N}^{k}\right\rangle =e^{-E_{N,N}^{k}\alpha_{t}}\left\vert
\Psi_{N,N}^{k}\right\rangle .
\end{equation}
Let us define $c_{N,N}^{k}$ as the inner product with the initial free fermion
ground state$,$%
\begin{equation}
c_{N,N}^{k}=\left\langle \Psi_{N,N}^{k\text{ }}\right.  \left\vert \Psi
_{N,N}^{0,\text{free}}\right\rangle .
\end{equation}
In the following we assume that $c_{N,N}^{0}$, the overlap between free
fermion ground state and the interacting ground state, is nonzero.

Let us define a transient energy expectation value that depends on the
Euclidean time $t,$
\begin{equation}
E_{N,N}(t)=\frac{1}{\alpha_{t}}\ln\frac{Z_{N,N}(t-\alpha_{t})}{Z_{N,N}(t)}.
\label{E_nn_t}%
\end{equation}
The spectral decomposition of $Z_{N,N}(t)$ gives%
\begin{equation}
Z_{N,N}(t)=\sum_{k}\left\vert c_{N,N}^{k}\right\vert ^{2}e^{-E_{N,N}^{k}t},
\end{equation}
and at large Euclidean time $t$ significant contributions\ come from only low
energy eigenstates. \ For large $t$ we find%
\begin{equation}
E_{N,N}(t)\approx E_{N,N}^{0}+%
{\displaystyle\sum\limits_{k\neq0}}
\frac{\left\vert c_{N,N}^{k}\right\vert ^{2}}{\left\vert c_{N,N}%
^{0}\right\vert ^{2}}\frac{e^{\left(  E_{N,N}^{k}-E_{N,N}^{0}\right)
\alpha_{t}}-1}{\alpha_{t}}e^{-\left(  E_{N,N}^{k}-E_{N,N}^{0}\right)  t}.
\label{transient_energy}%
\end{equation}
For low energy excitations $E_{N,N}^{k}-E_{N,N}^{0}$ is much smaller than the
energy cutoff scale $\alpha_{t}^{-1}$ imposed by the temporal lattice spacing.
\ Therefore
\begin{equation}
E_{N,N}(t)\approx E_{N,N}^{0}+%
{\displaystyle\sum\limits_{k\neq0}}
\frac{\left\vert c_{N,N}^{k}\right\vert ^{2}}{\left\vert c_{N,N}%
^{0}\right\vert ^{2}}\left(  E_{N,N}^{k}-E_{N,N}^{0}\right)  e^{-\left(
E_{N,N}^{k}-E_{N,N}^{0}\right)  t}. \label{transient_energy_2}%
\end{equation}
The ground state energy $E_{N,N}^{0}$ is given by the limit%
\begin{equation}
E_{N,N}^{0}=\lim_{t\rightarrow\infty}E_{N,N}(t). \label{ground_state}%
\end{equation}

\section{Importance sampling}

We calculate the ratio%
\begin{equation}
\frac{Z_{N,N}(t-\alpha_{t})}{Z_{N,N}(t)}%
\end{equation}
using Markov chain Monte Carlo. \ For $t=L_{t}\alpha_{t}$, configurations for
the auxiliary field are sampled according to the weight%
\begin{equation}
\exp\left\{  -U_{j}(s)+2\ln\left[  \left\vert \det\mathbf{M}_{j}(s,L_{t}%
\alpha_{t})\right\vert \right]  \right\}  ,
\end{equation}
where%
\begin{equation}
U_{1}(s)=U_{2}(s)=\frac{1}{2}\sum_{\vec{n},n_{t}}\left[  s(\vec{n}%
,n_{t})\right]  ^{2},
\end{equation}
and%
\begin{equation}
U_{3}(s)=U_{4}(s)=0.
\end{equation}

For $j=1,2,4$ importance sampling is implemented using hybrid Monte Carlo
\cite{Duane:1987de}. \ This involves computing molecular dynamics trajectories
for%
\begin{equation}
H_{j}(s,p)=\frac{1}{2}\sum_{\vec{n},n_{t}}\left[  p(\vec{n},n_{t})\right]
^{2}+V_{j}(s), \label{HMC_potential}%
\end{equation}
where $p(\vec{n},n_{t})$ is the conjugate momentum for $s(\vec{n},n_{t})$ and%
\begin{equation}
V_{j}(s)=U_{j}(s)-2\ln\left[  \left\vert \det\mathbf{M}_{j}(s,L_{t}\alpha
_{t})\right\vert \right]  .
\end{equation}
The steps of the algorithm are as follows.

\begin{itemize}
\item[Step 1:] Select an arbitrary initial configuration $s^{0}$.

\item[Step 2:] Select a configuration $p^{0}$ according to the Gaussian random
distribution%
\begin{equation}
P\left[  p^{0}(\vec{n},n_{t})\right]  \propto\exp\left\{  -\frac{1}{2}\left[
p^{0}(\vec{n},n_{t})\right]  ^{2}\right\}  .
\end{equation}

\item[Step 3:] For each $\vec{n},n_{t}$ let%
\begin{equation}
\tilde{p}^{0}(\vec{n},n_{t})=p^{0}(\vec{n},n_{t})-\frac{\varepsilon
_{\text{step}}}{2}\left[  \frac{\partial V_{j}(s)}{\partial s(\vec{n},n_{t}%
)}\right]  _{s=s^{0}} \label{step3}%
\end{equation}
for some small positive $\varepsilon_{\text{step}}$. \ The derivative of
$V_{j}$ is computed using%
\begin{align}
\frac{\partial V_{j}(s)}{\partial s(\vec{n},n_{t})}  &  =\frac{\partial
U_{j}(s)}{\partial s(\vec{n},n_{t})}-\frac{2}{\det\mathbf{M}_{j}}\sum
_{k,l}\frac{\partial\det\mathbf{M}_{j}}{\partial\left[  \mathbf{M}_{j}\right]
_{kl}}\frac{\partial\left[  \mathbf{M}_{j}\right]  _{kl}}{\partial s(\vec
{n},n_{t})}\nonumber\\
&  =\frac{\partial U_{j}(s)}{\partial s(\vec{n},n_{t})}-2\sum_{k,l}\left[
\mathbf{M}_{j}^{-1}\right]  _{lk}\frac{\partial\left[  \mathbf{M}_{j}\right]
_{kl}}{\partial s(\vec{n},n_{t})}.
\end{align}

\item[Step 4:] For steps $i=0,1,...,N_{\text{step}}-1$, let%
\begin{align}
s^{i+1}(\vec{n},n_{t})  &  =s^{i}(\vec{n},n_{t})+\varepsilon_{\text{step}%
}\tilde{p}^{i}(\vec{n},n_{t}),\\
\tilde{p}^{i+1}(\vec{n},n_{t})  &  =\tilde{p}^{i}(\vec{n},n_{t})-\varepsilon
_{\text{step}}\left[  \frac{\partial V_{j}(s)}{\partial s(\vec{n},n_{t}%
)}\right]  _{s=s^{i+1}},
\end{align}
for each $\vec{n},n_{t}.$

\item[Step 5:] For each $\vec{n},n_{t}$ let%
\begin{equation}
p^{N_{\text{step}}}(\vec{n},n_{t})=\tilde{p}^{N_{\text{step}}}(\vec{n}%
,n_{t})+\frac{\varepsilon_{\text{step}}}{2}\left[  \frac{\partial
V(s)}{\partial s(\vec{n},n_{t})}\right]  _{s=s^{N_{\text{step}}}}.
\end{equation}

\item[Step 6:] Select a random number $r\in$ $[0,1).$ \ If
\begin{equation}
r<\exp\left[  -H(s^{N_{\text{step}}},p^{N_{\text{step}}})+H(s^{0}%
,p^{0})\right]
\end{equation}
then set $s^{0}=s^{N_{\text{step}}}$. \ Otherwise leave $s^{0}$ as is. \ In
either case go back to Step 2 to start a new trajectory.
\end{itemize}

For $j=3$, hybrid Monte Carlo cannot be used since the auxiliary field has
only discrete values $-1$ or $+1$. \ In this case we use a local Metropolis
accept/reject update. \ For each sweep through the entire lattice a small
fraction of random lattice sites are selected. \ A new configuration
$s^{\prime}$ is produced by flipping the sign of $s$ at these sites. \ A
random number $r\in$ $[0,1)$ is selected and the new configuration $s^{\prime
}$ is accepted or rejected based on the Metropolis condition,%
\begin{equation}
r<\frac{\exp\left\{  -U_{3}(s^{\prime})+2\ln\left[  \left\vert \det
\mathbf{M}_{3}(s^{\prime},L_{t}\alpha_{t})\right\vert \right]  \right\}
}{\exp\left\{  -U_{3}(s)+2\ln\left[  \left\vert \det\mathbf{M}_{3}%
(s,L_{t}\alpha_{t})\right\vert \right]  \right\}  }.
\end{equation}

For all cases $j=1,2,3,4$ the observable we calculate for each configuration
is%
\begin{equation}
O_{j}(s,L_{t}\alpha_{t})=\frac{\left[  \det\mathbf{M}_{j}(s,\left(
L_{t}-1\right)  \alpha_{t})\right]  ^{2}}{\left[  \det\mathbf{M}_{j}%
(s,L_{t}\alpha_{t})\right]  ^{2}}.
\end{equation}
By taking the ensemble average of $O_{j}(s,L_{t}\alpha_{t})$ we obtain%
\begin{equation}
\frac{Z_{N,N}(t-\alpha_{t})}{Z_{N,N}(t)}%
\end{equation}
for $t=L_{t}\alpha_{t}$.

\section{Precision tests and performance comparisons}

\subsection{Precision tests for two particles}

We perform precision tests of the four auxiliary-field methods using the
two-particle system at unitarity. \ For the system with one up spin and one
down spin, we compute the dimensionless combination $mL^{2}E_{1,1}(t)$ for
$L=4$ and $L_{t}=6,12$. \ As explained in the previous section we use hybrid
Monte Carlo for $j=1,2,4$ and the Metropolis algorithm for $j=3$. The
simulations are performed using 16 processors starting with different random
number seeds producing independent configurations.\ \ The final result is
computed from the average of the individual processor results, and the
standard deviation is used to estimate the error of the average. \ The linear
space of states for the two-particle system is small enough that the transfer
matrix without auxiliary field in Eq.~(\ref{transfer_noaux}) can be used to
calculate exact results for comparison. \ The results are shown in Table
\ref{transfer_11}. \ For each case, $j=1,2,3,4$, the auxiliary-field Monte
Carlo results using $M_{j}$ reproduce the exact results up to the estimated
stochastic errors.

\begin{table}[tb]
\caption{Comparison of exact transfer matrix results and $j=1,2,3,4$
auxiliary-field Monte Carlo results for $mL^{2}E_{1,1}(t)$ with $L=4$ and
$L_{t}=6,12$.}%
\label{transfer_11}%
$%
\begin{tabular}
[c]{|c|c|c|c|c|c|}\hline
$L_{t}$ & Exact result & $M_{1}$ & $M_{2}$ & $M_{3}$ & $M_{4}$\\\hline
$6$ & $-2.087$ & $-2.06(3)$ & $-2.10(3)$ & $-2.05(4)$ & $-2.10(3)$\\\hline
$12$ & $-2.817$ & $-2.77(3)$ & $-2.81(4)$ & $-2.80(4)$ & $-2.88(4)$\\\hline
\end{tabular}
$\end{table}

\subsection{Performance comparisons}

In our discussion $E_{N,N}^{0,\text{free}}$ refers to the ground state energy
for $N$ up-spin and $N$ down-spin free fermions on the lattice. \ More
explicitly $E_{N,N}^{0,\text{free}}$ is the energy exponent of the
free-particle transfer matrix,%
\begin{equation}
:\exp\left(  -H_{\text{free}}\alpha_{t}\right)  :\left\vert \Psi
_{N,N}^{0,\text{free}}\right\rangle =\exp\left(  -E_{N,N}^{0,\text{free}%
}\alpha_{t}\right)  \left\vert \Psi_{N,N}^{0,\text{free}}\right\rangle ,
\end{equation}
for the same lattice parameters used to calculate $E_{N,N}(t)$. \ In contrast
with this lattice energy definition for $E_{N,N}^{0,\text{free}}$, we define
the Fermi energy $E_{F}$ purely in terms of particle density. \ For $N$ up
spins and $N$ down spins in a periodic cube we let%
\begin{equation}
E_{F}=\frac{k_{F}^{2}}{2m}=\frac{1}{2m}\left(  6\pi^{2}\frac{N}{L^{3}}\right)
^{2/3}\simeq7.596\frac{N^{2/3}}{mL^{2}}. \label{E_F}%
\end{equation}
For $N>1$ it is useful to define the dimensionless function%
\begin{equation}
\xi_{N,N}(t)=\frac{E_{N,N}(t)}{E_{N,N}^{0,\text{free}}}. \label{xsi_NN}%
\end{equation}
Scale invariance at unitarity requires that in the continuum limit $\xi
_{N,N}(t)$ depends only on the dimensionless combination $\frac{t}{mL^{2}}$.
\ For fixed $N$ the Fermi energy $E_{F}$ is proportional to $\frac{1}{mL^{2}}%
$, and so we can regard $\xi_{N,N}(t)$ as a universal function of $E_{F}t$.
\ This universal behavior provides a simple but nontrivial check of scale
invariance in our lattice calculations.

As one part of our analysis of computational performance we monitor the
rejection rate at Step 6 of the hybrid Monte Carlo algorithm for $j=1,2,4$ as
well as the Metropolis update for $j=3$. \ The average likelihood of rejection
is recorded as a rejection probability $P_{\text{r}}$. \ Another aspect we
monitor is the frequency of generating configurations with nearly singular
matrices $\mathbf{M}_{j}(s,L_{t}\alpha_{t})$. \ For this we introduce a small
positive parameter $\delta$ and reject configurations with%
\begin{equation}
\left\vert \det\mathbf{M}_{j}(s,L_{t}\alpha_{t})\right\vert <\delta
^{N}\left\vert \prod\limits_{i=1,...,N}\left[  \mathbf{M}_{j}(s,L_{t}%
\alpha_{t})\right]  _{ii}\right\vert \text{.} \label{singular}%
\end{equation}
By taking the limit $\delta\rightarrow0^{+}$ we can determine if
poorly-conditioned matrices make any detectable contribution to the final
result. \ The rejection rate due to these singular matrices is included in the
total rejection probability $P_{\text{r}}$, but is also recorded separately as
a singular matrix probability $P_{\text{s}}$.

We use double precision arithmetic in all calculations. \ Numerical
stabilization methods based on Gram-Schmidt orthogonalization have been
developed for determinantal quantum Monte Carlo simulations
\cite{Sugiyama:1986, Sorella:1989a, White:1989a}. \ These methods have proved
successful in reducing round-off error for singular matrix determinants and
should also be effective here. \ However in our ground state lattice
simulations we find that round-off error is only one of several problems
associated with singular matrix configurations. \ Another problem that arises
is quasi-non-ergodic behavior in the Monte Carlo updates. \ For the case of
hybrid Monte Carlo importance sampling this problem can be visualized in terms
of the landscape of the function $H_{j}(s,p)$ defined in
Eq.~(\ref{HMC_potential}). \ Each singular matrix configuration $s$ produces a
sharp peak in $H_{j}(s,p)$. \ Since hybrid Monte Carlo trajectories follow the
contour lines of $H_{j}(s,p)$, these trajectories can get trapped on orbits
around singular matrix configurations. \ Since the matrix inverse of
$\mathbf{M}_{j}$ diverges near these singular configurations, this behavior is
also accompanied by large fluctuations in the difermion spatial correlation
function
\begin{equation}
\left\langle a_{\downarrow}^{\dag}(\vec{n})a_{\uparrow}^{\dag}(\vec
{n})a_{\uparrow}(0)a_{\downarrow}(0)\right\rangle .
\end{equation}
This problem was observed in lattice simulations in \cite{Lee:2006hr}. \ For
these reasons we take a cautious approach to singular matrix configurations
until the problem is better understood. \ We use the singular matrix
probability $P_{\text{s}}$ as a diagnostic tool to measure the severity of the problem.

We use the unpolarized ten-particle system to test the computational
performance of the four auxiliary-field methods. \ We compute $\xi_{5,5}%
(L_{t}\alpha_{t})$ for spatial length $L=5$ and temporal lengths
$L_{t}=24,48,72$. \ The simulations are performed using 8 processors starting
with different random number seeds producing independent configurations.\ \ As
before the individual processor results are averaged, and the standard
deviation is used to estimate the error of the average. \ For $j=1,2,4$ we
generate hybrid Monte Carlo trajectories with $N_{\text{step}}=10$ and
$\varepsilon_{\text{step}}=0.1$. \ For the $j=3$ Metropolis update we flip the
sign of $s$ for $0.15\%$ of lattice sites selected randomly on each lattice
sweep. \ For the singular matrix condition in Eq.~(\ref{singular}), we use
$\delta=5\times10^{-7}$.

\begin{figure}
[ptb]
\begin{center}
\includegraphics[
height=2.7968in,
width=5.8064in
]%
{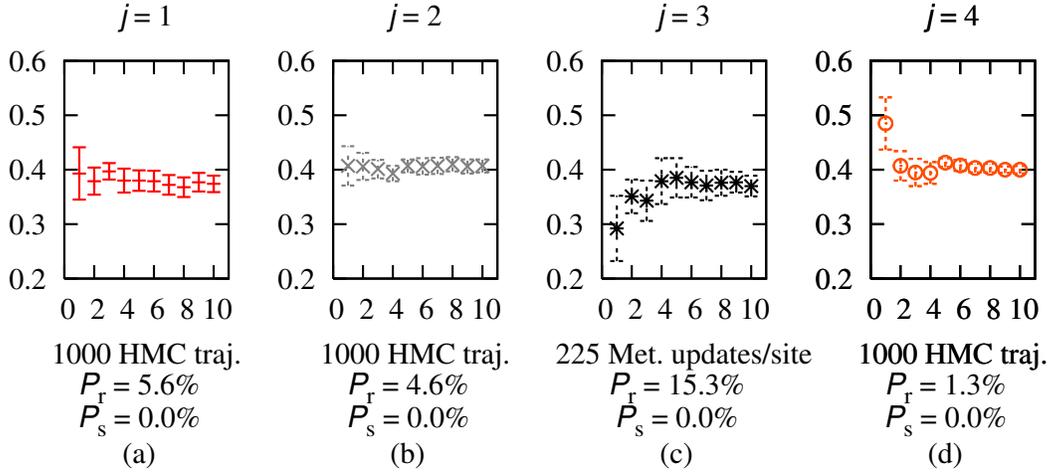}%
\caption{(Color online) Monte Carlo results for $\xi_{5,5}(L_{t}\alpha_{t})$
for $L=5$ and $L_{t}=24$. \ $P_{\text{r}}$ is the rejection probability, and
$P_{\text{s}}$ is the singular matrix probability.}%
\label{lt24}%
\end{center}
\end{figure}

Results for $\xi_{5,5}(L_{t}\alpha_{t})$ for $L_{t}=24$ are shown in
Fig.~\ref{lt24}. \ For $j=1,2,4$ the horizontal axis is the number of
completed hybrid Monte Carlo trajectories per processor. \ For $j=3$ the
horizontal axis is the number of attempted Metropolis updates per lattice
site. \ On an Intel Xeon processor the CPU run time for $1000$ hybrid Monte
Carlo trajectories is roughly the same as $150,000$ lattice sweeps with
Metropolis updating on $0.15\%$ of the sites. \ This corresponds with $225$
updates per site.%

All four calculations are in agreement with an average value $\xi
_{5,5}(24\alpha_{t})\approx0.39$. \ In all cases the estimated errors are
relatively small. \ In order of increasing error, the error bars for $j=4$ is
smallest, then $j=2$, $j=1$, and $j=3$. This is the same ordering we get by
sorting rejection probability $P_{\text{r}}$ from lowest to highest. \ The
singular matrix probability $P_{\text{s}}$ is zero in all cases.

\begin{figure}
[ptb]
\begin{center}
\includegraphics[
height=2.7959in,
width=5.8055in
]%
{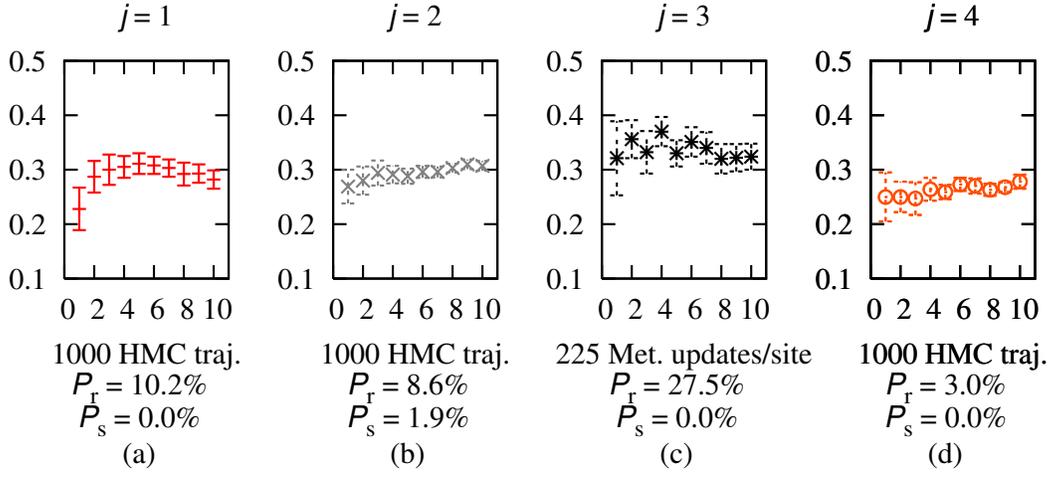}%
\caption{(Color online) Monte Carlo results for $\xi_{5,5}(L_{t}\alpha_{t})$
for $L=5$ and $L_{t}=48$.}%
\label{lt48}%
\end{center}
\end{figure}

Results for $L_{t}=48$ are shown in Fig.~\ref{lt48}.%
This time all four calculations are in agreement with $\xi_{5,5}(48\alpha
_{t})\approx0.29$. \ In all cases the estimated errors are still rather small,
but upon closer inspection there are some early signs of trouble for $j=3$ and
$j=2$. \ For $j=3$ the rejection probability is quite large. \ This suggests
difficulties in sampling the space of discrete auxiliary-field configurations.
\ For $j=2$ the singular matrix probability $P_{\text{s}}$ is $1.9\%$. \ This
is approaching the level where the contribution due to singular matrices may
be detectable above the background of stochastic error.

\begin{figure}
[ptb]
\begin{center}
\includegraphics[
height=2.7959in,
width=5.8055in
]%
{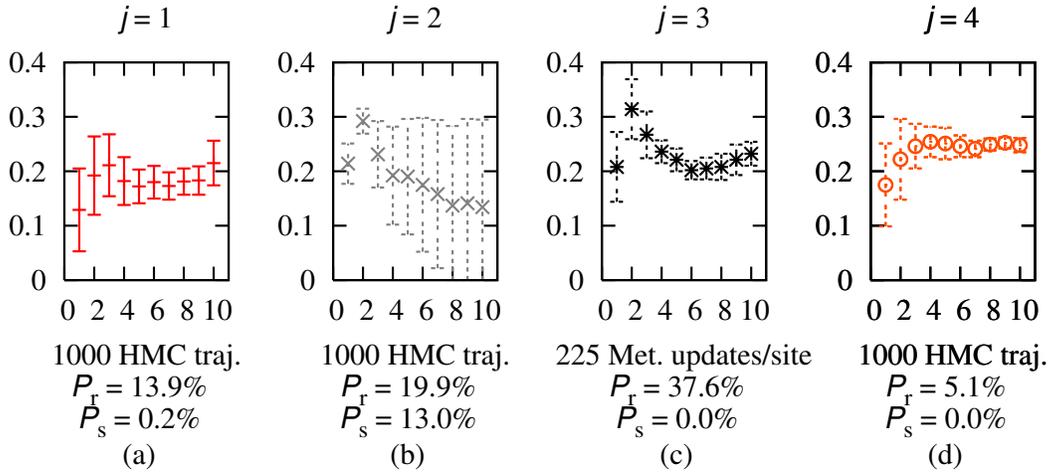}%
\caption{(Color online) Monte Carlo results for $\xi_{5,5}(L_{t}\alpha_{t})$
for $L=5$ and $L_{t}=72$.}%
\label{lt72}%
\end{center}
\end{figure}

Results for $L_{t}=72$ are shown in Fig.~\ref{lt72}.%
The simulation for $j=2$ has failed due to the high singular matrix
probability. \ The calculations for $j=1,3,4$ are consistent with a value of
$\xi_{5,5}(72\alpha_{t})\approx0.25$. \ However the rejection probability for
$j=3$ is very high. \ This could explain the gradual downward and then upward
drift in the $j=3$ data, a sign that the Monte Carlo simulation has not fully
equilibrated due to a long autocorrelation time. \ The same is true to a
lesser extent for $j=1$. \ For $j=1$ we also see that $P_{\text{s}}$ is small
but nonzero. \ This suggests that problems with singular matrices may appear
for somewhat larger $L$ and $L_{t}$.

From this analysis we rate the performance for $j=4$ superior to the other
three methods. \ This is not entirely unexpected. \ The use of a bounded
auxiliary field should reduce the likelihood of exceptional configurations
producing singular matrices. \ The use of a continuous auxiliary field makes
it possible to use hybrid Monte Carlo which is better at reducing rejection
probability than local Metropolis updates. \ For fixed trajectory length
$N_{\text{step}}\varepsilon_{\text{step}}$, the rejection probability for
hybrid\ Monte Carlo scales quadratically with step size, $\varepsilon
_{\text{step}}^{2}$. \ This is in contrast with the Metropolis algorithm,
where the rejection probability scales linearly with the fraction of lattice
sites updated in each sweep. \ This contributes to a much slower performance
of the Metropolis algorithm at large $L$ and $L_{t}$.

\section{Main results}

We use the bounded continuous auxiliary-field formulation $j=4$ for the
lattice results presented in this section. \ We consider the unpolarized
ten-particle and the fourteen-particle systems. \ For the calculation of
$\xi_{5,5}(L_{t}\alpha_{t})$ we use the lattice dimensions $L^{3}\times L_{t}$
shown in Table~\ref{dimensions_55}. \ For $\xi_{7,7}(L_{t}\alpha_{t})$ we use
the lattice dimensions $L^{3}\times L_{t}$ in Table~\ref{dimensions_77}.
\ \begin{table}[tb]
\caption{Lattice dimensions $L^{3}\times L_{t}$ used in calculations for
$\xi_{5,5}(L_{t}\alpha_{t}).$}%
\label{dimensions_55}%
\begin{tabular}
[c]{|c|c|c|c|c|c|}\hline
$L^{3}$ & $4^{3}$ & $5^{3}$ & $6^{3}$ & $7^{3}$ & $8^{3}$\\\hline
$L_{t}$ &
\begin{tabular}
[c]{c}%
$16$\\
$20$\\
$\vdots$\\
$48$%
\end{tabular}
&
\begin{tabular}
[c]{c}%
$24$\\
$30$\\
$\vdots$\\
$72$%
\end{tabular}
&
\begin{tabular}
[c]{c}%
$36$\\
$42$\\
$\vdots$\\
$102$%
\end{tabular}
&
\begin{tabular}
[c]{c}%
$48$\\
$56$\\
$\vdots$\\
$144$%
\end{tabular}
&
\begin{tabular}
[c]{c}%
$70$\\
$80$\\
$\vdots$\\
$190$%
\end{tabular}
\\\hline
\end{tabular}
\end{table}\begin{table}[tbtb]
\caption{Lattice dimensions $L^{3}\times L_{t}$ used in calculations for
$\xi_{7,7}(L_{t}\alpha_{t}).$}%
\label{dimensions_77}%
$%
\begin{tabular}
[c]{|c|c|c|c|c|c|}\hline
$L^{3}$ & $4^{3}$ & $5^{3}$ & $6^{3}$ & $7^{3}$ & $8^{3}$\\\hline
$L_{t}$ &
\begin{tabular}
[c]{c}%
$12$\\
$16$\\
$\vdots$\\
$44$%
\end{tabular}
&
\begin{tabular}
[c]{c}%
$18$\\
$24$\\
$\vdots$\\
$72$%
\end{tabular}
&
\begin{tabular}
[c]{c}%
$30$\\
$36$\\
$\vdots$\\
$96$%
\end{tabular}
&
\begin{tabular}
[c]{c}%
$40$\\
$48$\\
$\vdots$\\
$128$%
\end{tabular}
&
\begin{tabular}
[c]{c}%
$50$\\
$80$\\
$\vdots$\\
$170$%
\end{tabular}
\\\hline
\end{tabular}
$\end{table}We use trajectory parameters $N_{\text{step}}=10$, $\varepsilon
_{\text{step}}=0.1,$ and singular matrix parameter $\delta=5\times10^{-7}$.
\ The rejection probability $P_{\text{r}}$ reaches a maximum of $12\%$ and the
singular matrix probability $P_{\text{s}}$ reaches a maximum of $3\%$ for the
largest lattice volume, $L^{3}=8^{3}$ and $L_{t}\geq170$. \ For most of the
lattice simulations the values for $P_{\text{r}}$ and $P_{\text{s}}$ are much
smaller. \ The simulations are run with a minimum of $16$ processors each
running a minimum of $10,000$ hybrid Monte Carlo trajectories.

\bigskip From Eq.~(\ref{transient_energy_2}) we expect%
\begin{equation}
\xi_{N,N}(t)\approx\xi_{N,N}+%
{\displaystyle\sum\limits_{k\neq0}}
\frac{\left\vert c_{N,N}^{k}\right\vert ^{2}}{\left\vert c_{N,N}%
^{0}\right\vert ^{2}}\left(  \frac{E_{N,N}^{k}-E_{N,N}^{0}}{E_{N,N}^{0}%
}\right)  e^{-\left(  E_{N,N}^{k}-E_{N,N}^{0}\right)  t} \label{transient_xsi}%
\end{equation}
at large $t$. \ To extract $\xi_{N,N}$ we perform a least squares fit of
$\xi_{N,N}(t)$ to the functional form,%
\begin{equation}
\xi_{N,N}(t)=\xi_{N,N}+b_{N,N}e^{-\delta_{N,N}E_{F}t}.
\end{equation}
For asymptotically large $E_{F}t$, we can identify $\delta_{N,N}E_{F}$ with
the energy separation between the ground state and the first excited state,
possibly with degenerate partners. \ For large $N$ the lowest excitation is
expected to be a two phonon state with zero total momentum. For large $N$ the
excitation energy for this state is small compared with $E_{F}$ and therefore
$\delta_{N,N}<<1$. \ 

Our observable $\xi_{N,N}(t)$ is proportional to the expectation value of the
energy. \ When the energy difference $E_{N,N}^{k}-E_{N,N}^{0}$ is very small
compared to $E_{N,N}^{0}$ the contribution proportional to $e^{-\left(
E_{N,N}^{k}-E_{N,N}^{0}\right)  t}$ is difficult to resolve against the
background of stochastic noise. \ We note the factor
\begin{equation}
\frac{E_{N,N}^{k}-E_{N,N}^{0}}{E_{N,N}^{0}}%
\end{equation}
multiplying $e^{-\left(  E_{N,N}^{k}-E_{N,N}^{0}\right)  t}$ in
Eq.~(\ref{transient_xsi}). \ If the objective were to measure $E_{N,N}%
^{k}-E_{N,N}^{0}$ accurately for very low excitations, then it would be more
effective to compute the Euclidean time projection of some other observable
such as the difermion spatial correlation function
\begin{equation}
\left\langle a_{\downarrow}^{\dag}(\vec{n})a_{\uparrow}^{\dag}(\vec
{n})a_{\uparrow}(0)a_{\downarrow}(0)\right\rangle .
\end{equation}
This technique was used to identify the lowest energy excitations for several
unpolarized lattice systems at unitarity \cite{Lee:2006hr}.

In the analysis here we focus only on measuring the ground state energy
accurately and ignore the numerically small contributions hidden in the
asymptotic tail of $\xi_{N,N}(t)$. \ We determine $b_{N,N},\delta_{N,N}%
,\xi_{N,N}$ from least squares fitting over the range $E_{F}t=2$ to $E_{F}%
t=9$. \ The values $b_{N,N}$ and $\delta_{N,N}$ we determine from least
squares fitting should be interpreted as a spectral average,
\begin{equation}
\sum_{k\neq0}\frac{\left\vert c_{N,N}^{k}\right\vert ^{2}}{\left\vert
c_{N,N}^{0}\right\vert ^{2}}\left(  \frac{E_{N,N}^{k}-E_{N,N}^{0}}{E_{N,N}%
^{0}}\right)  e^{-\left(  E_{N,N}^{k}-E_{N,N}^{0}\right)  t}\approx
b_{N,N}e^{-\delta_{N,N}E_{F}t}.
\end{equation}
Since $\xi_{N,N}(t)$ has a well-defined continuum limit for fixed $E_{F}t$,
each of the dimensionless parameters $b_{N,N},\delta_{N,N},\xi_{N,N}$ also has
a well-defined continuum limit. \ Up to uncertainties the size of least
squares fitting errors, $\xi_{N,N}$ is independent of the initial state
overlap amplitudes $c_{N,N}^{k}$. \ However $b_{N,N}$ and $\delta_{N,N}$ both
depend on $c_{N,N}^{k}$. \ Table~\ref{fit5_three} shows the three-parameter
fit results for $N=5,$ and Table~\ref{fit7_three} shows the three-parameter
fit results for $N=7$. \ \begin{table}[tb]
\caption{Results for the three-parameter fit for $\xi_{5,5}(t)$ using
$b_{5,5}$, $\delta_{5,5},\xi_{5,5}$.}%
\label{fit5_three}
$%
\begin{tabular}
[c]{|c|c|c|c|c|}\hline
$L$ & $b_{5,5}$ & $\delta_{5.5}$ & $\xi_{5,5}$ & $\chi^{2}$/d.f.\\\hline
$4$ & $0.42(12)$ & $0.47(13)$ & $0.22(2)$ & $1.2$\\\hline
$5$ & $0.40(7)$ & $0.44(7)$ & $0.23(1)$ & $0.4$\\\hline
$6$ & $0.41(8)$ & $0.42(8)$ & $0.24(1)$ & $1.6$\\\hline
$7$ & $0.81(18)$ & $0.69(10)$ & $0.26(1)$ & $0.6$\\\hline
$8$ & $0.36(12)$ & $0.32(15)$ & $0.24(4)$ & $0.6$\\\hline
\end{tabular}
$\end{table}\begin{table}[tbtb]
\caption{Results for the three-parameter fit for $\xi_{7,7}(t)$ using
$b_{7,7}$, $\delta_{7,7},\xi_{7,7}$.}%
\label{fit7_three}
$%
\begin{tabular}
[c]{|c|c|c|c|c|}\hline
$L$ & $b_{7,7}$ & $\delta_{7,7}$ & $\xi_{7,7}$ & $\chi^{2}$/d.f.\\\hline
$4$ & $0.24(9)$ & $0.39(13)$ & $0.26(1)$ & $1.3$\\\hline
$5$ & $0.28(4)$ & $0.35(6)$ & $0.27(1)$ & $1.0$\\\hline
$6$ & $0.39(7)$ & $0.45(6)$ & $0.29(1)$ & $1.3$\\\hline
$7$ & $0.29(8)$ & $0.27(15)$ & $0.26(4)$ & $1.7$\\\hline
$8$ & $0.36(7)$ & $0.41(10)$ & $0.30(2)$ & $0.8$\\\hline
\end{tabular}
$\end{table}The average chi-square per degree of freedom for the fits is about
$1$. \ The error estimates for the fit parameters are calculated by explicit
simulation. \ We introduce Gaussian-random noise scaled by the error bars of
each data point for $\xi_{N,N}(t)$. \ The fit is repeated many times with the
random noise included to estimate the one standard-deviation spread in the fit parameters.

The error in $\xi_{N,N}$ would be considerably smaller if the fit needed only
two parameters rather than three parameters. \ This can be arranged if we
neglect the $L$-dependence of $\delta_{N,N}$ and fix $\delta_{N,N}$ according
to the average values for $L=4,5,6,7,8$ as quoted in Tables~\ref{fit5_three}
and \ref{fit7_three}. \ Since $\delta_{N,N}$ has a well-defined continuum
limit, the $L$-dependence of $\delta_{N,N}$ should in fact be small. \ For
$N=5$ the average value is $\delta_{5,5}=0.47$, and for $N=7$ the average
value is $\delta_{7,7}=0.37$. \ Using these values we refit $\xi_{N,N}(t)$
with the two parameters $b_{N,N},\xi_{N,N}$. \ Table~\ref{fit5_two} shows the
two-parameter fit results for $N=5,$ and Table~\ref{fit7_two} shows the
two-parameter fit results for $N=7$.

\begin{table}[tb]
\caption{Results for the two-parameter fit for $\xi_{5,5}(t)$ using
$b_{5,5},\xi_{5,5}$ with $\delta_{5,5}=0.47.$}%
\label{fit5_two}
$%
\begin{tabular}
[c]{|c|c|c|c|}\hline
$L$ & $b_{5,5}$ & $\xi_{5,5}$ & $\chi^{2}$/d.f.\\\hline
$4$ & $0.42(4)$ & $0.223(5)$ & $1.0$\\\hline
$5$ & $0.43(4)$ & $0.236(2)$ & $0.3$\\\hline
$6$ & $0.47(3)$ & $0.247(2)$ & $1.5$\\\hline
$7$ & $0.51(4)$ & $0.242(5)$ & $0.9$\\\hline
$8$ & $0.47(4)$ & $0.264(6)$ & $0.6$\\\hline
\end{tabular}
$\end{table}

\begin{table}[tb]
\caption{Results for the two-parameter fit for $\xi_{7,7}(t)$ using
$b_{7,7},\xi_{7,7}$ with $\delta_{7,7}=0.37$.}%
\label{fit7_two}
$%
\begin{tabular}
[c]{|c|c|c|c|}\hline
$L$ & $b_{7,7}$ & $\xi_{7,7}$ & $\chi^{2}$/d.f.\\\hline
$4$ & $0.23(2)$ & $0.261(4)$ & $1.1$\\\hline
$5$ & $0.29(1)$ & $0.276(3)$ & $0.9$\\\hline
$6$ & $0.31(2)$ & $0.285(2)$ & $1.4$\\\hline
$7$ & $0.34(4)$ & $0.282(9)$ & $1.6$\\\hline
$8$ & $0.33(3)$ & $0.294(6)$ & $0.8$\\\hline
\end{tabular}
$\end{table}

\begin{figure}
[ptb]
\begin{center}
\includegraphics[
height=3.026in,
width=3.3875in
]%
{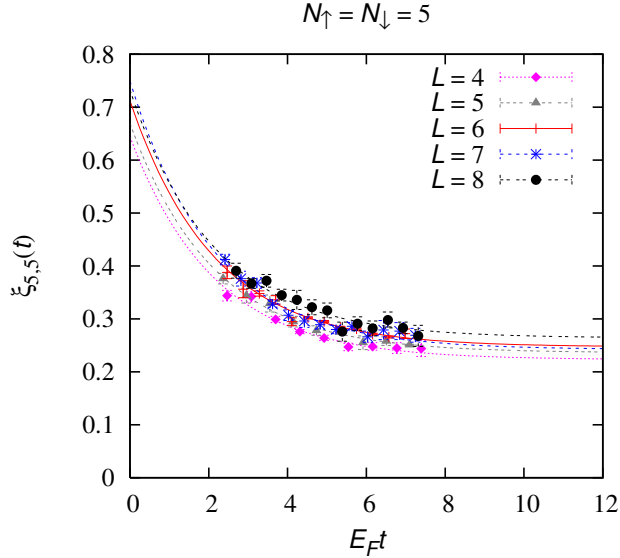}%
\caption{(Color online) The lattice data for $\xi_{5,5}(t)$ versus $E_{F}t$
for $L=4,5,6,7,8$. \ Also shown are the results of the two-parameter fits.}%
\label{n5}%
\end{center}
\end{figure}

\begin{figure}
[ptbptb]
\begin{center}
\includegraphics[
height=3.0251in,
width=3.3875in
]%
{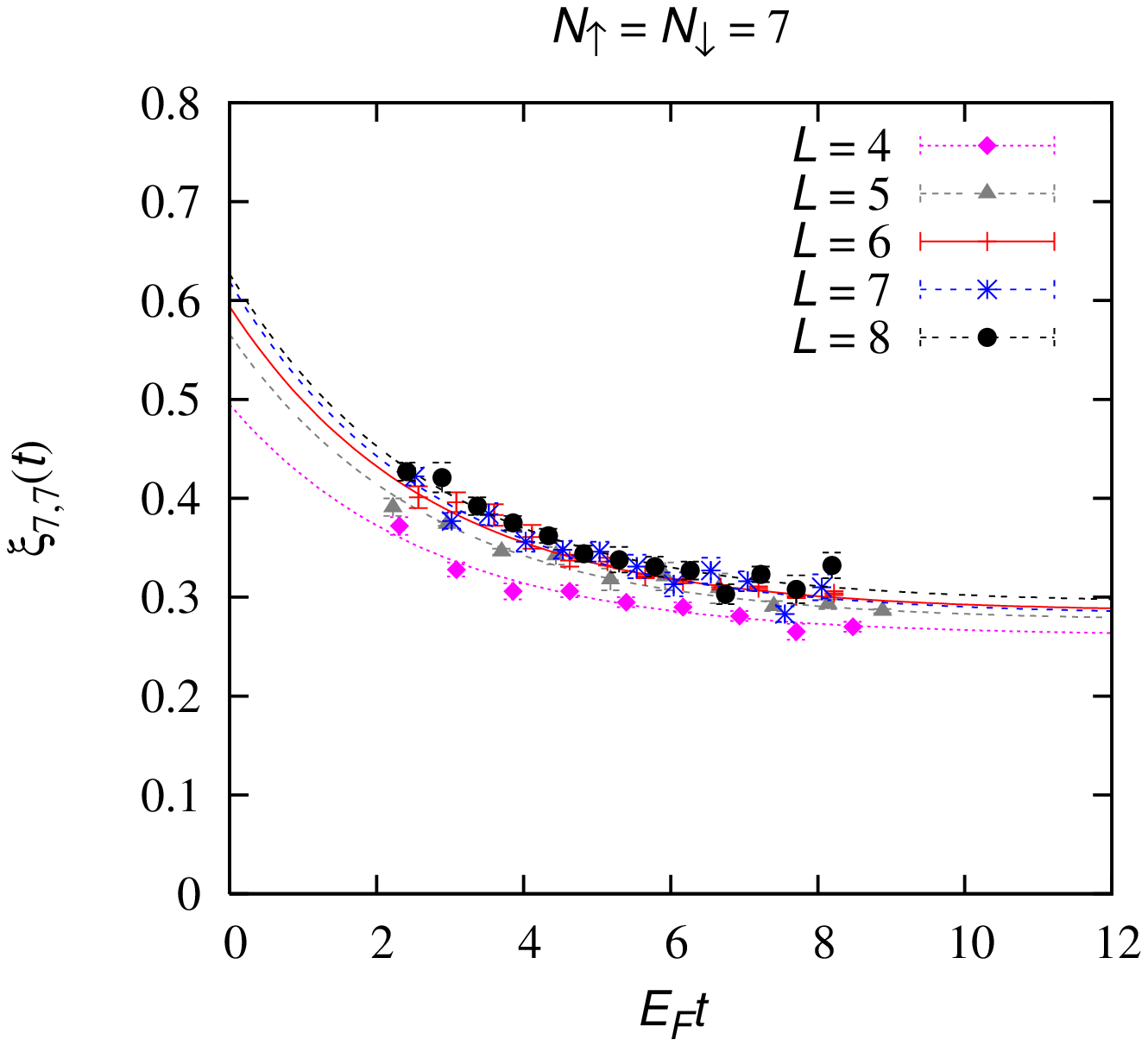}%
\caption{(Color online) The lattice data for $\xi_{7,7}(t)$ versus $E_{F}t$
for $L=4,5,6,7,8$. \ Also shown are the results of the two-parameter fits.}%
\label{n7}%
\end{center}
\end{figure}

The average chi-square per degree of freedom is again about $1$. \ The lattice
data for $\xi_{5,5}(t)$ together with the two-parameter fit functions are
shown Fig.~\ref{n5}.%
Fig.~\ref{n7} shows the lattice data for $\xi_{7,7}(t)$ with two-parameter fit
functions. \
The lattice results show that $\xi_{5,5}(t)$ and $\xi_{7,7}(t)$ are both
approximately universal functions of $E_{F}t$ independent of $L$. \ This was
expected from the scale invariance of the unitarity limit.

\begin{figure}
[ptb]
\begin{center}
\includegraphics[
height=2.7181in,
width=3.179in
]%
{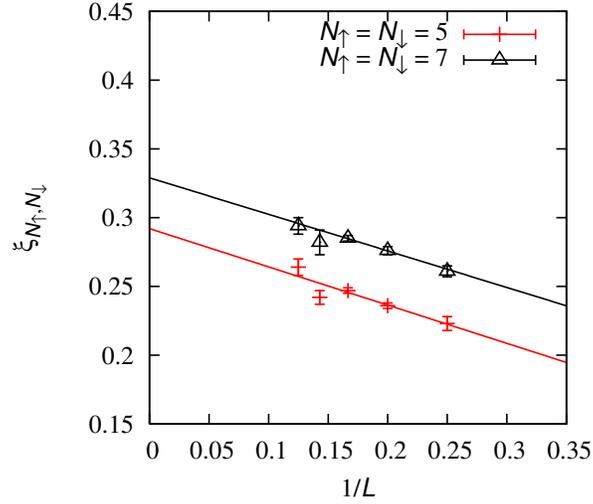}%
\caption{(Color online) Results from the two-parameter fits for $\xi_{5,5}$
and $\xi_{7,7}$ at finite $L$ and the corresponding continuum limit
extrapolations.}%
\label{ldependence}%
\end{center}
\end{figure}

Using the results for $\xi_{5,5}$ and $\xi_{7,7}$ for $L=4,5,6,7,8$ we can
extrapolate to the continuum limit $L\rightarrow\infty$. \ We expect some
residual dependence on $L$ proportional to $L^{-1},$ arising from effects such
as the effective range correction, broken Galilean invariance, and possibly
other lattice cutoff effects. \ From the three-parameter fit results in
Tables~\ref{fit5_three} and \ref{fit7_three}, the linear extrapolation in
$L^{-1}$ gives the continuum limit values%
\begin{equation}
\xi_{5,5}=0.308(25),
\end{equation}%
\begin{equation}
\xi_{7,7}=0.337(30).
\end{equation}
If we extrapolate the two-parameter fit results in Tables~\ref{fit5_two} and
\ref{fit7_two} we get%
\begin{equation}
\xi_{5,5}=0.292(12), \label{xsi_55}%
\end{equation}%
\begin{equation}
\xi_{7,7}=0.329(5). \label{xsi_77}%
\end{equation}
Results from the two-parameter fits for $\xi_{5,5}$ and $\xi_{7,7}$ at finite
$L$ and the corresponding continuum limit extrapolations are shown in
Fig.~\ref{ldependence}.%
We note that the continuum limit fit for $\xi_{N,N}$ could also be performed
with $E_{N,N}^{0,\text{free}}$ defined in the continuum limit for the same
cubic box size. \ This procedure is not recommended since it introduces a
larger $L^{-2}$ dependence and degrades the quality of the linear $L^{-1}$
fit. \ However the fit can be done and the extrapolated values for $\xi_{5,5}$
and $\xi_{7,7}$ are each about $0.015$ higher than the values reported in
Eq.~(\ref{xsi_55}) and (\ref{xsi_77}) with somewhat larger error bars.

\section{Discussion}

In \cite{Lee:2005fk} the values%
\begin{equation}
\xi_{5,5}=0.24(2),
\end{equation}%
\begin{equation}
\xi_{7,7}=0.28(2),
\end{equation}
were found based on an average of lattice results for $L=4,5,6$. \ The results
here for $\xi_{5,5}$ and $\xi_{7,7}$ with $L=4,5,6$ are consistent with these
values. \ The continuum limit extrapolation was not possible in
\cite{Lee:2005fk} due to problems with increasing singular matrix probability
$P_{\text{s}}$. \ That calculation used the auxiliary field formulation $j=2,$
which had the largest $P_{\text{s}}$ of the four methods considered here.
\ The bounded continuous auxiliary-field method appears to solve this problem
for the lattice systems considered here. \ The values for $\xi_{5,5}$ and
$\xi_{7,7}$ from small lattices $L=4,5,6$ are each shifted upwards by $0.05$
when extrapolated to the continuum limit.

In the calculations presented here we have only considered systems with $10$
and $14$ particles and have not attempted to determine the thermodynamical
limit $N\rightarrow\infty$. \ It is therefore interesting to compare with
results obtained using other methods that have computed ground state energies
for both small and large values of $N$. \ Each of these other methods contain
some unknown systematic errors, and so a benchmark comparison with continuum
extrapolated Monte Carlo lattice results for $N=5$ and $N=7$ provides an
estimate of the systematic error.

There is a discrepancy of about $0.13$ in the reported values for $\xi_{5,5}$
and $\xi_{7,7}$ between\ continuum extrapolated lattice results and fixed-node
Green's function Monte Carlo results on a periodic cube. \ The fixed-node
Green's function\ Monte Carlo simulations find\ $\xi_{N,N}=0.44(1)$ for $5\leq
N\leq21$ \cite{Carlson:2003z} and $0.42(1)$ for larger $N$
\cite{Astrakharchik:2004,Carlson:2005xy}. \ This discrepancy suggests that the
upper bound on the ground state energy using fixed-node Green's function Monte
Carlo might be lowered further by a more optimal fermionic nodal surface.

\begin{figure}
[ptb]
\begin{center}
\includegraphics[
height=2.7172in,
width=3.179in
]%
{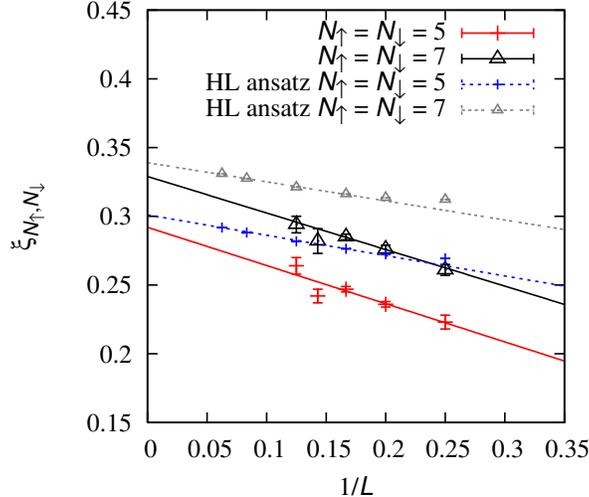}%
\caption{(Color online) Comparison of results from Fig.~\ref{ldependence} with
Hamiltonian lattice results using the symmetric heavy-light ansatz in the
lowest filling approximation.}%
\label{ldependence_hl}%
\end{center}
\end{figure}

A recent Hamiltonian lattice study computed $\xi_{N,N}$ for the range $N=2$ to
$N=32$ on cubic lattices up to $L^{3}=16^{3}$ \cite{Lee:2007A}. \ This
calculation used a method called the\ symmetric heavy-light ansatz in the
lowest filling approximation. \ In Fig.~\ref{ldependence_hl} we compare Monte
Carlo lattice results presented here and symmetric heavy-light ansatz results
for $N=5$ and $N=7$. \
The continuum limit extrapolations of the symmetric heavy-light results give%
\begin{equation}
\xi_{5,5}=0.301(1),
\end{equation}%
\begin{equation}
\xi_{7,7}=0.339(1).
\end{equation}
These results are within $0.01$ of the values found in Eq.~(\ref{xsi_55}) and
(\ref{xsi_77}). \ This level of accuracy is consistent with the size of errors
found in \cite{Lee:2007A} for four-body and six-body systems of the 1D, 2D, 3D
attractive Hubbard models at arbitrary coupling.

The different $L^{-1}$ slopes for the Hamiltonian lattice and Euclidean
lattice extrapolations in Fig.~\ref{ldependence_hl} are consistent with the
fact that the effective range for the Hamiltonian lattice interaction is a
smaller negative fraction of the lattice spacing than for the Euclidean
lattice transfer matrix. \ However there are other effects such as broken
Galilean invariance on the lattice which produce a similar $L^{-1}$ dependence
\cite{Lee:2007jd}. \ An accurate calculation of the effective range correction
with controlled systematic errors requires either an effective range larger
than the lattice spacing or an analysis of the effect of changing the
effective range parameter relative to the lattice spacing. \ The effective
range correction has recently been computed on the lattice with realistic
dilute neutron matter at next-to-leading order in chiral effective field
theory \cite{Borasoy:2007vy,Borasoy:2007vk}.

\begin{figure}
[ptb]
\begin{center}
\includegraphics[
height=3.5258in,
width=4.1909in
]%
{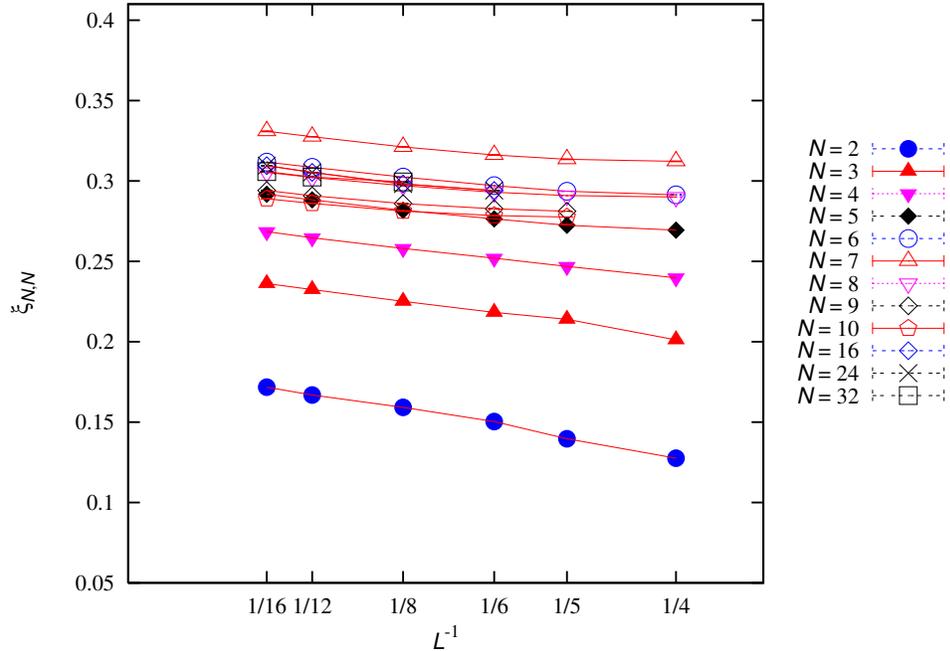}%
\caption{(Color online) Results for $\xi_{N,N}$ at unitarity using the
symmetric heavy-light ansatz in the lowest filling approximation
\cite{Lee:2007A}.}%
\label{ldependence_markov}%
\end{center}
\end{figure}

Results from the symmetric heavy-light ansatz for general $N$ are shown in
Fig.~\ref{ldependence_markov} \cite{Lee:2007A}. \ We note that the value for
$\xi_{N,N}$ reaches a maximum at $N=7$. \ This can be explained by the closed
shell at $N=7$ in the free fermion ground state and the absence of shell
effects in the interacting system. \ The numerical agreement for the benchmark
comparisons at $N=5$ and $N=7$ provides some confidence in the symmetric
heavy-light ansatz value of $\xi=0.31(1)$ for the unitarity limit in the
continuum and thermodynamic limits \cite{Lee:2007A}.%

The bounded continuous auxiliary-field method succeeds in reducing the
singular matrix probability $P_{\text{s}}$ by minimizing large fluctuations in
the auxiliary-field transfer matrix elements. \ Very roughly this corresponds
with reducing the size of%
\begin{equation}
\max_{s\left(  \vec{n},n_{t}\right)  }\left\vert A_{j}\left[  s\left(  \vec
{n},n_{t}\right)  \right]  \right\vert ,
\end{equation}
given the constraints%
\begin{equation}
\int d_{j}s\left(  \vec{n},n_{t}\right)  \,1=1,
\end{equation}%
\begin{equation}
\int d_{j}s\left(  \vec{n},n_{t}\right)  \,A_{j}\left[  s\left(  \vec{n}%
,n_{t}\right)  \right]  =0,
\end{equation}%
\begin{equation}
\int d_{j}s\left(  \vec{n},n_{t}\right)  \,A_{j}^{2}\left[  s\left(  \vec
{n},n_{t}\right)  \right]  =-C\alpha_{t}.
\end{equation}
For lattice systems larger than the ones considered here the problem with
singular matrices will reappear. \ For very large systems there may be no
choice but to use the stabilization methods developed in \cite{Sugiyama:1986,
Sorella:1989a, White:1989a} to reduce round-off error and confront the
problems of large fluctuations and quasi-non-ergodic behavior using
brute-force large-scale simulations. \ However for moderately larger systems
it may be possible to gain further advantage using a function $A_{j}\left[
s(\vec{n},n_{t})\right]  $ with a steep slope at $s(\vec{n},n_{t})=0$ and
relatively flat away from zero. \ The most extreme case would be to use an odd
periodic step function. \ But this is equivalent to the discrete
auxiliary-field formulation $j=3$ with the problems discussed concerning large
rejection probabilities. \ However with a smooth approximation to an odd
periodic step function, it may be possible to reduce the singular matrix
probability while at the same time compensate for the increase in rejection
probability with a smaller hybrid Monte Carlo step size, $\varepsilon
_{\text{step}}.$

\section{Summary}

We have presented new Euclidean lattice methods which remove some
computational barriers encountered in previous lattice calculations of the
ground state energy in the unitarity limit. \ We compared the performance of
four different auxiliary-field methods that produce exactly the same lattice
transfer matrix. \ By far the best performance was obtained using a bounded
continuous auxiliary field with hybrid Monte Carlo updating. \ With this
method we calculated results for $10$ and $14$ fermions at lattice volumes
$4^{3},5^{3},6^{3},7^{3},8^{3}$ and extrapolated to the continuum limit. \ For
$10$ fermions in a periodic cube, we found the ground state energy to be
$0.292(12)$ times the ground state energy for non-interacting fermions. \ For
$14$ fermions the ratio is $0.329(5)$. \ These values may be useful as
benchmarks for calculations of the unitarity limit ground state using other methods.

\section*{Acknowledgements}

The author is grateful for discussions with Cliff Chafin, Gautam Rupak, and
Thomas Sch\"{a}fer. This work is supported in part by DOE grant DE-FG02-03ER41260.

\bibliographystyle{apsrev}
\bibliography{NuclearMatter}

\end{document}